\documentclass[prl,twocolumn]{revtex4}

\usepackage{amsmath,amssymb,bbm,graphicx}

\begin{document}


\title{Deterministic Secure Direct Communication Using Entanglement}

\author{Kim Bostr\" om, Timo Felbinger}
\affiliation{Institut f\"ur Physik, Universit\"at Potsdam, 
14469 Potsdam, Germany
}


\begin{abstract}

A novel secure communication protocol is presented, based on an entangled pair of qubits and allowing asymptotically secure key distribution and quasi-secure direct communication. Since the information is transferred in a deterministic manner, no qubits have to be discarded. The transmission of information is instantaneous, i.e. the information can be decoded during the transmission. The security against arbitrary eavesdropping attacks is provided. In case of eavesdropping attacks with full information gain, the detection rate is 50\% per control transmission. 
The experimental realization of the protocol is feasible with relatively small effort, which also makes commercial applications conceivable.  

\end{abstract}

\maketitle

\textit{Introduction.---} 
Cryptographic schemes based on quantum mechanics are usually \emph{non-deterministic} \cite{BB84,Ekert,Bruss}. Alice, the sender, can encode a classical bit into a quantum state, which is then sent to Bob, but she cannot determine the bit value that Bob will finally \emph{decode}. Inspite of that, such non-deterministic communication can be used to establish a \emph{shared secret key} between Alice and Bob, consisting of a sequence of random bits.
This secret key can then be used to encrypt a message which is sent through a classical public channel. Recently, a novel quantum communication protocol has been presented \cite{Almut1} that allows \emph{secure direct communication}, where the message is deterministically send through the quantum channel, but can only be decoded after a final transmission of classical information.
We present a communication scheme, the ``ping-pong protocol'', that also allows for deterministic communication. This protocol can be used for the transmission of either a secret key or a plaintext message. In the latter case, the protocol is \emph{quasi-secure}, i.e. an eavesdropper is able to gain a small amount of message information before being detected. In case of a \emph{key transmission} the protocol is asymptotically secure.
In contrast to other quantum cryptographic schemes, the presented scheme is \emph{instantaneous}, i.e. the information can be decoded during the transmission and no final transmission of additional information is needed. 
The basic idea of the protocol, encoding information by local operations on an EPR pair, has already been raised by Bennett and Wiesner \cite{BW92}. In our protocol, we follow this idea, but abandon the \emph{dense} coding feature in favour of a \emph{secure} transmission.

\begin{figure}[t]
	\[{\includegraphics[width=0.4\textwidth]{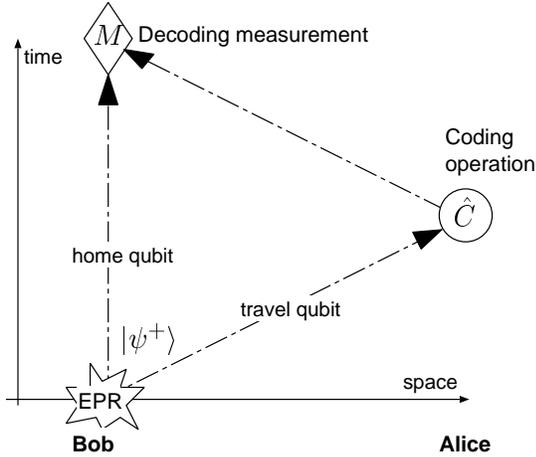}}\]
	\vspace*{-0.6cm}\caption{\small Message mode. Dashed lines are qubit transfers.}\label{ppmessage}
\end{figure}

\begin{figure}
	\[{\includegraphics[width=0.4\textwidth]{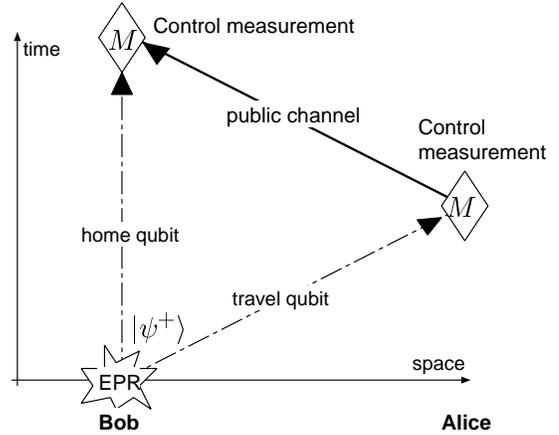}}\]
	\vspace*{-0.6cm}\caption{\small Control mode. Solid lines are classical transfers.}\label{ppcontrol}
\end{figure}

\textit{The ping-pong protocol.---} 
When two photons are maximally \emph{entangled} in their polarization degree of freedom, then each single photon is not polarized at all. Denote the horizontal and vertical polarization state by $|0\rangle$ and $|1\rangle$, respectively, then the \emph{Bell states}
$|\psi^\pm\rangle=\frac1{\sqrt2}(|01\rangle\pm|10\rangle)$
are maximally entangled states in the two-particle Hilbert space ${\cal H}={\cal H}_A\otimes{\cal H}_B$. A measurement of the polarization of one photon, say $A$, leads to a completely random result. This is reflected by the fact that the  corresponding \emph{reduced density matrices},
$\rho^\pm_A:={\rm Tr}_B\{|\psi^\pm\rangle\langle\psi^\pm|\}$
are both equal to the complete mixture, $\rho^\pm_A=\frac12{\mathbbm1}_A$.
Hence, no experiment performed on only one photon can distinguish these states from each other.
However, since the states $|\psi^\pm\rangle$ are mutually orthogonal, a measurement on \emph{both} photons can perfectly distinguish the states from each other. In other words: One bit of information can be encoded in the states $|\psi^\pm\rangle$, which is completely unavailable to anyone who has only access to one of the photons. As one can easily verify, the unitary operator
$\hat\sigma_z^{A}\equiv(\hat\sigma_z\otimes{\mathbbm1})
	=(|0\rangle\langle0|-|1\rangle\langle1|)
	\otimes{\mathbbm1}$
flips between the two states $|\psi^\pm\rangle$,
\begin{equation}
	\hat\sigma_z^A|\psi^\pm\rangle=|\psi^\mp\rangle.
\end{equation}
Altough $\hat\sigma_z^A$ acts \emph{locally}, i.e. on one photon only, it has a \emph{non-local} effect. Someone who has access to one single photon only, can \emph{encode} one bit of information, but he cannot \emph{decode} it, since he has no access to the other photon.
This is a situation perfectly suited for a cryptographic scenario. 
Bob prepares two photons in the state $|\psi^+\rangle$. He keeps one photon, the ``\emph{home qubit}'', and sends the other one, the ``\emph{travel qubit}'', to Alice (``ping!''). 
Alice decides either to perform the operation $\hat\sigma_z$ on the travel qubit or to do nothing, i.e. to perform the operation ${\mathbbm1}$.
Then she sends the travel qubit back to Bob (``pong!'').
Bob, who has now both qubits again, performs a Bell measurement resulting in either $|\psi^+\rangle$ or $|\psi^-\rangle$, depending on what Alice did. Thus, he has received one bit of information from Alice.
One qubit travels forth and back (``ping-pong!'') and one bit of information flows from Alice to Bob.
Let us introduce two communication modes, ``\emph{message mode}'' and ``\emph{control mode}'' (see Figs.~\ref{ppmessage},\ref{ppcontrol}). By default, Alice and Bob are in message mode and communicate the way described above. With probability $c$, Alice switches to control mode and instead of performing her operation on the travel qubit, she performs a measurement in the basis ${\cal B}_z=\{|0\rangle,|1\rangle\}$. Using the public channel, she sends the result to Bob, who then also switches to control mode and performs a measurement in the same basis ${\cal B}_z$. Bob compares his own result with Alice's result. If both results coincide, Bob knows that Eve is in the line and stops the communication.
[t]
Let us give an explicit algorithm for the protocol.
\begin{enumerate}
\renewcommand{\labelenumi}{p.\arabic{enumi})}
\setcounter{enumi}{-1}
\item\label{pinit}
Protocol is initialized. $n=0$.
The message to be transmitted is a sequence $x^N=(x_1,\ldots,x_N)$, where $x_n\in\{0,1\}$.
\item\label{pstart}
$n=n+1$. Alice and Bob are set to message mode.
Bob prepares two qubits in the Bell state $|\psi^+\rangle=\frac1{\sqrt2}(|01\rangle+|10\rangle)$.
\item\label{pstart2}
He stores one qubit, the \emph{home qubit}, and sends the other one, the \emph{travel qubit}, to Alice through the quantum channel.
\item
Alice receives the travel qubit. With probability $c$ she switches to control mode and proceeds with~c.1, else she proceeds with m.1.
\begin{enumerate}\renewcommand{\labelenumii}{c.\arabic{enumii})}
	\item\label{cstart}
	Alice measures the travel qubit in the basis ${\cal B}_z$ and obtains the result $i\in\{0,1\}$ with equal probability.
	\item
	She sends $i$ through the public channel to Bob.
	\item
	Bob receives $i$ from the public channel, switches to control mode and measures the home qubit in the basis ${\cal B}_z$ resulting in the value $j$.
	\item
	($i=j$): Eve is detected. Abort transmission. 
	($i\neq j$): Set $n=n-1$ and Goto~p.\ref{pstart}.
\end{enumerate}
\begin{enumerate}\renewcommand{\labelenumii}{m.\arabic{enumii})}
	\item\label{mstart}
	Define $\hat C_0:={\mathbbm1}$ and $\hat C_1:=\hat\sigma_z$.
	For $x_n\in\{0,1\}$, Alice performs the coding operation $\hat C_{x_n}$ on the travel qubit and sends it back to Bob.
	\item
	Bob receives the travel qubit and performs a Bell measurement on both qubits resulting in the final state $|\psi'\rangle\in\{|\psi^+\rangle,|\psi^-\rangle\}$. He decodes the message as
	\begin{equation}
		|\psi'\rangle=\begin{cases}
			|\psi^+\rangle\Rightarrow x_n=0\\
			|\psi^-\rangle\Rightarrow x_n=1
			\end{cases}.
	\end{equation}
	\item
	($n<N$): Goto~p.\ref{pstart}.
	($n=N$): Goto~p.\ref{pfinal}.
\end{enumerate}
\item\label{pfinal}
Message $x^N$ is transmitted from Alice to Bob. Communication successfully terminated.
\end{enumerate} 

\begin{figure}[b]
	\[{\includegraphics[width=0.4\textwidth]{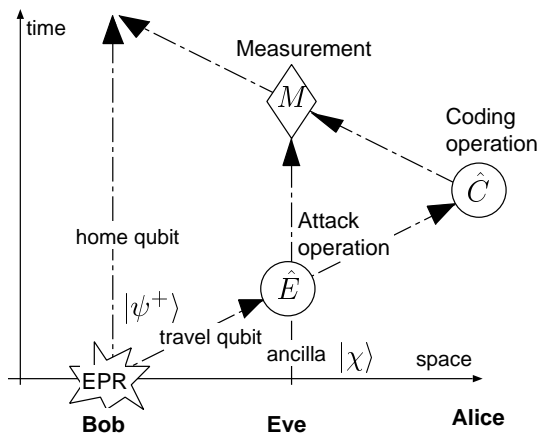}}\]
	\vspace*{-0.6cm}\caption{\small A general eavesdropping attack.}\label{attack}
\end{figure}

\textit{Security proof.---}
Eve is an evil quantum physicist able to build all devices that are allowed by the laws of quantum mechanics.
Her aim is to find out which operation Alice performs.
Eve has no access to Bob's home qubit, so all her operations are restricted to the travel qubit, whose state is (to Eve) indistinguishable from the complete mixture
$\rho_A={\rm Tr}_B\{|\psi^+\rangle\langle\psi^+|\}=\frac12{\mathbbm1}_A$.
The most general quantum operation is a \emph{completely positive map} ${\cal E}:{\cal S}({\cal H}_A)\rightarrow{\cal S}({\cal H}_A)$ on the state space ${\cal S}({\cal H}_A)$. Due to the \emph{Stinespring dilation theorem} \cite{Stinespring}, any completely positive map can be realized by a unitary operation on a larger Hilbert space. For ${\cal H}_A$ and ${\cal E}$ given, there is an \emph{ancilla space} ${\cal H}_E$ of dimension $\dim{\cal H}_E\leq(\dim{\cal H}_A)^2$, an ancilla state $|\chi\rangle\in{\cal H}_E$, and a unitary operation $\hat E$ on ${\cal H}_A\otimes{\cal H}_E$, such that for all states $\rho_A\in{\cal S}({\cal H}_A)$, we have
\begin{equation}
	{\cal E}(\rho_A)={\rm Tr}_E\{\hat E(\rho_A\otimes|\chi\rangle\langle\chi|)
	\hat E^\dagger\}.
\end{equation}
In order to gain information about Alice's operation, Eve should first perform the unitary \emph{attack operation} $\hat E$ on the composed system, then let Alice perform her coding operation $\hat C$ on the travel qubit, and finally perform a measurement on the composed system (see Fig.~\ref{attack}).
Since a probable control measurement by Alice takes place \emph{before} Eve's final measurement, the latter has no influence on the detection probability for Eve's attack. All that can be detected is the attack operation $\hat E$.
Let us analyze the detection probability $d$, given an attack operation $\hat E$. Since for Eve the state of the travel qubit is indistinguishable from the complete mixture, we can replace the state of the travel qubit by the \emph{a priori} mixture $\rho_A=\frac12|0\rangle\langle0|+\frac12|1\rangle\langle 1|$,
which corresponds to the situation where Bob sends the travel qubit in either of the states $|0\rangle$ or $|1\rangle$, with equal probability $p=1/2$.
Let us at first consider the case where Bob sends $|0\rangle$.
Alice adds an ancilla in the state $|\chi\rangle$ and performs the unitary operation $\hat E$ on both systems, resulting in
\begin{eqnarray}
	|\psi'\rangle&=&\hat E|0,\chi\rangle
		=\alpha|0,\chi_0\rangle+\beta|1,\chi_1\rangle,
\end{eqnarray}
where $|\chi_0\rangle,|\chi_1\rangle$ are pure ancilla states uniquely determined by $\hat E$, and $|\alpha|^2+|\beta|^2=1$.
In a subsequent control measurement, Alice measures the travel qubit in the basis ${\cal B}_z=\{|0\rangle,|1\rangle\}$ and sends the result to Bob. Without Eve, the result will always read ``0'', hence the detection probability for Eve's attack in a control run reads
\begin{equation}\label{d}
	d=|\beta|^2=1-|\alpha|^2.
\end{equation}
Now let us analize how much information Eve can maximally gain when there is no control run. 
After Eve's attack operation, the state of the system reads
\begin{widetext}
\begin{eqnarray}
	\rho'&=&|\psi'\rangle\langle\psi'|
		=|\alpha|^2|0,\chi_0\rangle\langle0,\chi_0|
			+|\beta|^2|1,\chi_1\rangle\langle1,\chi_1|
			+\alpha\beta^*|0,\chi_0\rangle\langle1,\chi_1|
			+\alpha^*\beta|1,\chi_1\rangle\langle0,\chi_0|,
\end{eqnarray}
\end{widetext}
which can be rewritten in the orthogonal basis $\{|0,\chi_0\rangle,|1,\chi_1\rangle\}$ as
\begin{equation}
	\rho'=\begin{pmatrix}|\alpha|^2&\alpha\beta^*\\\alpha^*\beta&|\beta|^2
		\end{pmatrix}.
\end{equation}
Alice encodes her bit by applying the operation $\hat C_0={\mathbbm1}$ or $\hat C_1=\hat\sigma_z$ to the travel qubit, with probability $p_0$ and $p_1$, respectively. The state of the travel qubit after Eve's attack operation and after Alice's encoding operation reads
\begin{equation}
	\rho''=\begin{pmatrix}|\alpha|^2&\alpha\beta^*(p_0-p_1)\\
	\alpha^*\beta(p_0-p_1)&|\beta|^2
		\end{pmatrix}.
\end{equation}
The maximal amount $I_0$ of classical information that can be extracted from this state is given by the \emph{von-Neumannn entropy},
$I_0=S(\rho'')\equiv -{\rm Tr}\{\rho''\log_2\rho''\}$.
In order to calculate the von-Neumann entropy we need the eigenvalues $\lambda$ of $\rho''$, which are the roots of the characteristic polynomial
$\det(\rho''-\lambda{\mathbbm1})$, yielding the two eigenvalues
\begin{equation}
	\lambda_{1,2}=\frac12\left(1\pm\sqrt{1-4|\alpha\beta|^2[1-(p_0-p_1)^2]}
	\right),
\end{equation}
so we have
\begin{equation}\label{I0lambda}
	I_0=-\lambda_1\log_2\lambda_1-\lambda_2\log_2\lambda_2.
\end{equation}
The maximal information gain $I_0$ can be expressed as a function of the detection probability $d$.
Using~(\ref{d}), we have
$|\alpha\beta|^2=(1-|\beta|^2)|\beta|^2=(d-d^2)$,
and therefore
\begin{equation}\label{lambdad}
	\lambda_{1,2}=\frac12\pm\frac12\sqrt{1-(4d-4d^2)[1-(p_0-p_1)^2]}.
\end{equation}
Now assume that Bob sends $|1\rangle$ rather than $|0\rangle$. The above calculations can be done in full analogy, resulting in the same crucial relations~(\ref{I0lambda},\ref{lambdad}). Eve's task is, of course, to minimize $d$. Though if she chooses an eavesdropping action $\hat E$ that provides $d=0$, then $\lambda_1=1,\,\lambda_2=0$, which implies $I_0=0$, therefore Eve can gain no information at all.
Thus we have shown:

\textit{Any effective eavesdropping attack can be detected.---}
In the case $p_0=p_1=1/2$, where Alice encodes exactly 1 bit, expression~(\ref{lambdad}) simplifies to
$\lambda_{1,2}=\frac12\pm|\frac12-d|$,
or $\lambda_1=d,\,\lambda_2=1-d$. Interestingly, the maximal information gain is equal to the Shannon entropy of a binary channel,
\begin{equation}\label{I0}
	I_0(d)= -d\log_2d-(1-d)\log_2(1-d).
\end{equation}
The function $I_0(d)$ has a maximum at $d=1/2$, and can be inversed on the interval $[0,1/2]$, giving a monotonous function $0\leq d(I_0)\leq 1/2,\,I_0\in[0,1]$.
By choosing a desired information gain $I_0>0$ per attack, Eve has to face a detection probability $d(I_0)>0$. If she wants to gain the \emph{full} information ($I_0=1$), the detection probability is $d(I_0=1)=1/2$.

\textit{Direct communication versus key distribution.---}
In contrast to quantum key distribution protocols like BB84 \cite{BB84}, the ping-pong protocol provides a \emph{deterministic} transmission of bits, hence it is possible to communicate the message \emph{directly} from Alice to Bob. Assuming that Eve wants to gain full information in each attack, the ping-pong protocol provides a detection probability of $d=1/2$, which is significantly higher than the detection probability of the BB84 protocol, where we have $d=\frac12\times\frac12=\frac14$ for the same situation.
Furthermore, the BB84 protocol has a probability of $1/2$ that a transmitted bit has to be discarded due to the wrong choice of basis on both sides. 

\begin{figure}[t]
	\[{\includegraphics[width=0.4\textwidth]{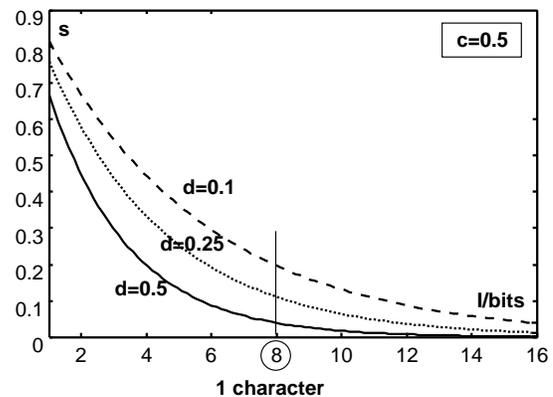}}\]
	\vspace*{-0.6cm}\caption{\small Eavesdropping success probability as a function of the maximal eavesdropped information, plotted for different detection probabilities $d$.}\label{success}
\end{figure}

Taking into account the probability $c$ of a control run, the \emph{effective transmission rate}, i.e. the number of message bits per protocol run, reads $r=1-c$,
which is equal to the probability for a message transfer.
Say, Eve wants to eavesdrop one message transfer without being detected. 
The probability for this event reads
\begin{eqnarray}
	s(c,d)&=&(1-c)+c(1-d)(1-c)+\nonumber \\
		&&c^2(1-d)^2(1-c)+\ldots\\
		&=&\frac{1-c}{1-c(1-d)},
\end{eqnarray}
where the terms in the (geometric) series correspond to Eve having to survive $0,1,2,\ldots$ control runs before she gets to eavesdrop on a message run, finally yielding the desired information of $I_0(d)$ bits. 
After $n$ successful attacks Eve gains $nI_0(d)$ bits of information and survives with probability $s^n$, thus the probability to successfully eavesdrop $I=nI_0(d)$ bits reads
$s(I,c,d)=s(c,d)^{I/I_0(d)}$, so
\begin{equation}
	s(I,c,d)=\left(\frac{1-c}{1-c(1-d)}\right)^{I/I_0(d)},
\end{equation}
where $I_0(d)$ is given by~(\ref{I0}).	
For $c>0$, $d>0$, this value decreases exponentially but is nonzero. In the limit $I\rightarrow\infty$ (a message or key of infinite length) we have $s\rightarrow0$, so the protocol is \emph{asymptotically secure}, just like the BB84 protocol.
Let us give an example. A convenient choice of the control parameter is $c=0.5$, where on the average every second bit is a control bit.
Say, Eve wants to gain full information in each attack, thus $I_0=1$ and $d=1/2$.  The probability that Eve successfully eavesdrops 1 character (8 bits) is already as low as $s\approx0.039$. 
In Fig.~\ref{success} we have plotted the eavesdropping success probability as a function of the information gain $I$, for $c=0.5$ and for different detection probabilities $d$ that Eve can choose. (Note that for $d<1/2$ Eve only gets \emph{part} of the message right and does not even know \emph{which} part.)
If desired, the security can arbitrarily be improved by increasing the control parameter $c$ at the cost of decreasing the transmission rate. 
Let us call such communication ``\emph{quasi-secure}''. If we want a \emph{perfectly} secure communication (which is, strictly speaking, also not really perfect), we must abandon the direct transfer in favour of a \emph{key} transfer. In this case, Alice does not transmit the message directly to Bob but rather takes a random sequence of $N$ bits from a secret random number generator. After a succesful transmission, the random sequence is used as a \emph{shared secret key} between Alice and Bob.
Eve has virtually no advantage in eavesdropping only a few bits, because one can choose classical privacy amplification protocols that make it \emph{very} hard to decode parts of the message with only \emph{some} of the key bits given. The one-time-pad scheme, by the way, is not quite a good choice, because here each eavesdropped key bit directly yields one decoded message bit. 
Anyway, as soon as Eve is detected, the transfer stops and she has learned nothing but a sequence of nonsense random bits. 

\textit{Experimental feasibility.---}
The Bell state $|\psi^+\rangle$ can be created by \emph{parametric down-conversion}. Bob's Bell measurement must only distinguish between the states $|\psi^\pm\rangle$, which can be accomplished, too. The storage of one photon is necessary only for a duration corresponding to twice the distance between Alice and Bob. 
The encoding procedure corresponds to a controlled $\hat\sigma_z$-operation, which can be realized by triggered optical elements. The correlation test involves a simple measurement of the linear polarization in a fixed basis.
Altogether, the experimental realization of the ping-pong protocol should be feasible using nowaday's technology. Even a commercial application could be envisaged.

We had fruitful discussions with Almut Beige, Luke Rallan, Jens Eisert, Martin Plenio, Sougato Bose, and others. 
This work is supported by the Deutsche Forschungsgemeinschaft (DFG) and by the European Union (EQUIP).


\end{document}